\documentclass[12pt,a4paper]{article}
\usepackage{a4wide}
\usepackage{latexsym}
\usepackage{epsf}
\usepackage{amssymb}

\begin{document}

\begin{flushright}
\small
IFT-UAM/CSIC-02-36\\
{\bf hep-th/0209069}\\
September $8$th, $2002$
\normalsize
\end{flushright}

\begin{center}


\vspace{2cm}

{\Large {\bf The Near-Horizon Limit of the Extreme Rotating $d=5$
    Black Hole as a Homogenous Spacetime}}

\vspace{2cm}

{\bf\large Natxo Alonso-Alberca}${}^{\spadesuit\heartsuit}$
\footnote{E-mail: {\tt natxo@leonidas.imaff.csic.es}},
{\bf\large Ernesto Lozano-Tellechea}${}^{\spadesuit\heartsuit}$
\footnote{E-mail: {\tt Ernesto.Lozano@uam.es}}\\
{\bf\large and Tom{\'a}s Ort\'{\i}n}${}^{\spadesuit\clubsuit}$
\footnote{E-mail: {\tt Tomas.Ortin@cern.ch}}

\vspace{1cm}

${}^{\spadesuit}$\ {\it Instituto de F\'{\i}sica Te{\'o}rica, C-XVI,
Universidad Aut{\'o}noma de Madrid\\
Cantoblanco, E-28049 Madrid, Spain}

\vskip 0.2cm
${}^{\heartsuit}$\ {\it Departamento de F\'{\i}sica Te{\'o}rica, C-XI,
Universidad Aut{\'o}noma de Madrid\\
Cantoblanco, E-28049 Madrid, Spain}

\vskip 0.2cm
${}^{\clubsuit}$\ {\it I.M.A.F.F., C.S.I.C., 
Calle de Serrano 113 bis\\ 
E-28006 Madrid, Spain}
\vspace{.7cm}


{\bf Abstract}

\end{center}

\begin{quotation}

\small

We show that the spacetime of the near-horizon limit of the extreme
rotating $d=5$ black hole, which is maximally supersymmetric in
$N=2,d=5$ supergravity for any value of the rotation parameter $j\in
[-1,1]$, is locally isomorphic to a homogeneous non-symmetric
spacetime corresponding to an element of the 1-parameter family of
coset spaces $\frac{SO(2,1)\times SO(3)}{SO(2)_{j}}$ in which the
subgroup $SO(2)_{j}$ is a combination of the two $SO(2)$ subgroups of
$SO(2,1)$ and $SO(3)$.

\end{quotation}

\newpage

\pagestyle{plain}


\section*{Introduction}

The vast majority of the known maximally supersymmetric solutions of
supergravity theories seem to be symmetric spaces: Minkowski or $AdS$
spacetimes, products of $AdS$ spacetimes and spheres $AdS_{m}\times
S^{n}$ or H$pp$-wave spacetimes. Their Killing vectors and spinors and
their relations that determine their supersymmetry algebras can be
found by simple geometrical methods \cite{Alonso-Alberca:2002gh}.

The only exception seems to be the near-horizon limit of the extreme rotating
$d=5$ black holes
\cite{Cvetic:1998xh,Gauntlett:1998fz,Gibbons:1999uv,Herdeiro:2000ap}.  This
solution can be written in the form \cite{Lozano-Tellechea:2002pn}

\begin{equation}
\label{eq:solution}
\left\{
  \begin{array}{rcl}
ds^{2} & = & R^{2}d\Pi_{(2)}^{2} -R^{2}d\Omega_{(2)}^{2}
-R^{2}(d\psi +\cos{\alpha} \cos{\theta} d\varphi -\sin{\alpha}\, {\rm sinh} 
\chi d\phi)^{2}\, ,\\
& & \\
F & = & \sqrt{3}R\cos{\alpha}\, {\rm cosh} \chi d\chi\wedge d\phi
-\sqrt{3}R\sin{\alpha}\sin{\theta}\, d\theta\wedge d\varphi\, ,\\
  \end{array}
\right.  
\end{equation}

\noindent
where 

\begin{equation}
  \begin{array}{rcl}
d\Pi_{(2)}^{2} & = & {\rm cosh}^{2}\chi\, d\phi^{2} -d\chi^{2}  \, ,\\
& & \\
d\Omega_{(2)}^{2} & = & d\theta^{2}+\sin^{2}{\theta}\,d\varphi^{2}\, ,\\
  \end{array}
\end{equation}

\noindent
are respectively the metrics of the unit radius $AdS_{2}$ spacetime
and the unit radius 2-sphere $S^{2}$. The rotation parameter $j$ is
here $\cos{\alpha}$.

The metric of this solution looks like a sort of twisted product
$AdS_{3}\times S^{3}$ in which the sphere and the $AdS$ spacetime
share a common direction parametrized by $\psi$. Actually, when
$\cos{\alpha}=1$ (the purely electric solution), the dimension $\psi$
belongs only to the sphere and the metric is exactly that of
$AdS_{2}\times S^{3}$ and, when $\cos{\alpha}=0$, the dimension $\psi$
belongs entirely to the $AdS$ spacetime and the metric is exactly that
$AdS_{3}\times S^{2}$. These are singular limits, though, because the
isometry group is the 7-dimensional $SO(2,1)\times SO(3)\times SO(2)$
for generic values of $\cos{\alpha}$ but becomes the 9-dimensional
$SO(2,1)\times SO4)$ or $SO(2,2)\times SO(3)$ in the two limits

Not surprisingly, the solution can be obtained by dimensional
reduction of the $AdS_{3}\times S^{3}$ solution of $N=2,d=6$
supergravity along a direction which is a linear combination of the
two $S^{1}$ fibers of the Hopf fibrations
$AdS_{3}\stackrel{S^{1}}{\rightarrow}AdS_{2}$ and
$S^{3}\stackrel{S^{1}}{\rightarrow}S^{2}$
\cite{Lozano-Tellechea:2002pn}.  It can also be obtained by
dimensional oxidation of the dyonic Robinson-Bertotti solution
\cite{kn:Rob,kn:Bert} of $N=2,d=4$ supergravity
\cite{Lozano-Tellechea:2002pn}, (whose metric is that of
$AdS_{2}\times S^{2}$ and is also maximally supersymmetric
\cite{Gibbons:1984kp,Kallosh:1992gu}) and these dimensional relations
give us very important clues about the geometry of the solution and
how to find a coset construction of its metric \cite{Coquereaux:ne}.

In fact, these relations immediately suggest that the metric could be
constructed as an invariant metric over the coset $\frac{SO(2,1)\times
  SO(3)}{SO(2)_{j}}$, in which the subgroup $SO(2)_{j}$ is a combination of he
two $SO(2)$ subgroups of $SO(2,1)$ and $SO(3)$, that is: the group manifold
$SO(2,1)$ equipped with the bi-invariant metric can be identified with the
$AdS_{3}$ spacetime and the coset $SO(2,1)/SO(2)$ with the left-invariant
metric can be identified (locally) with the $AdS_{2}$ spacetime. Analogously,
the group manifold $SO(3)$ equipped with the bi-invariant metric can be
identified (locally) with the $S^{3}$ spacetime and the coset $SO(3)/SO(2)$
with the left-invariant metric can be identified with the $S^{2}$ spacetime.
In the product $AdS_{3}\times S^{3}$ there are two $SO(2)$ subgroups available
for taking the quotient (which is equivalent to dimensional reduction) and one
choice gives, in $d=5$ $AdS_{2}\times S^{3}$ and the other $AdS_{3}\times
S^{2}$. One could also take the quotient over the $SO(2)_{j}$ subgroup
generated by a linear combination of the generators of the two above-mentioned
$SO(2)$ subgroups and the left-invariant metric should be the one in
Eq.~(\ref{eq:solution}).

There is another $SO(2)$ subgroup present, generated by the orthogonal
linear combination. This $SO(2)$ commutes with the other one and
belongs to its normalizer, which is $SO(2)\times SO(2)$. It is a
well-known fact \cite{Coquereaux:ne} that the isometry group of the
left-invariant metric over a coset $G/H$ is, generically $G\times
N(H)/H$, where $N(H)$ is the normalizer of $H$ and $N(H)/H$ is the
right isometry group. Here $N(H)/H=SO(2)_{j}$ and then the full isometry
group should be the 7-dimensional $SO(2,1)\times SO(3)\times SO(2)$,
as we want. In the two singular limits, there is enhancement of the
isometry group as explained above.

In this paper we are going to prove that our proposal is indeed correct by
explicitly constructing first the metric in Eq.~(\ref{eq:solution}) as a
left-invariant metric over the coset\footnote{Our identification of the
  near-horizon limit of the rotating extreme black hole and the coset space is
  only local. We will not be concerned with global issues here.}
$\frac{SO(2,1)\times SO(3)}{SO(2)_{j}}$.  The spacetime, is, thus,
homogeneous, but it is not symmetric. Secondly, we are then going to use this
construction to find the Killing vectors and spinors, although we will find
difficulties to relate them, due to the fact that in our construction we will
not use the Killing metric, but instead we will use the Minkowski metric,
which is also $SO(2)$-invariant: the Killing metric of the real form
$so(2,1)\times so(3)$ has the signature $(--+---)$, i.e.~the $so(2,1)$ part
has the wrong signature in our conventions (mostly minus signature), but this
can not be corrected by means of analytic continuation (one gets complex
metrics or metrics with wrong signature). Fortunately, the Minkowski metric
has the necessary properties.


\section*{Construction of the Metric and Killing Vectors}

The Lie algebra of $SO(2,1)$ can be written in the form

\begin{equation}
[T_{i},T_{j}]=-\epsilon_{ijk}\mathsf{Q}^{kl}T_{l}\, ,
\,\,\,\,
i,j,\cdots=1,2,3,\, ,  
\hspace{1cm}
\mathsf{Q}={\rm diag}\, (++-)\, , 
\end{equation}

\noindent
and its Killing metric is $K=2{\rm diag}\, (++-)$. To construct
$AdS_{2}$, one has to take the coset $SO(2,1)/SO(2)$ where the
subgroup $SO(2)$ is generated indistinctly by $T_{1}$ or $T_{2}$. We
will choose for the sake of definiteness $T_{1}$. The projection of
the Killing metric on the orthogonal subspace generated by
$T_{2},T_{3}$ ${\rm diag}\, (+-)$ has the right signature to give
$AdS_{2}$. Actually, the signature is the opposite to our mostly minus
conventions, but a global factor is immaterial and the time
coordinate, compact, is associated to $T_{3}$ (the $-$ sign in the
Killing metric).

It is important to observe that there is no real form of this algebra
with Killing metric $K={\rm diag}\, (--+)$. Also, we are forced to
associate the time coordinate with $T_{3}$.
 
The Lie algebra of $SO(3)$ can be written in the form

\begin{equation}
[\tilde{T}_{i},\tilde{T}_{j}]=-\epsilon_{ijk}\tilde{T}_{k}\, ,
\,\,\,\,
i,j,\cdots=1,2,3,\, ,   
\end{equation}

\noindent
and its Killing metric is $K=2{\rm diag}\, (---)$. To construct
$S^{2}$, one has to take the coset $SO(3)/SO(2)$ where the subgroup
$SO(2)$ is generated by any of the generators $\tilde{T}_{i}$. We will
choose $T_{3}$ for definiteness. Observe that there is no real form
with Killing metric $K=2{\rm diag}\, (+++)$.

The subgroup $SO(2)$ that we will use will be the one generated by the
combination

\begin{equation}
M \equiv \cos{\alpha} T_{1} +\sin{\alpha}\tilde{T}_{3}\, .  
\end{equation}

We now make the following redefinitions

\begin{equation}
P_{0}={\textstyle\frac{1}{R}}T_{3}\, ,
\,\,\,\,
P_{1}={\textstyle\frac{1}{R}}T_{2}\, ,
\,\,\,\,
P_{2}={\textstyle\frac{1}{R}}\tilde{T}_{1}\, ,
\,\,\,\,
P_{3}={\textstyle\frac{1}{R}}\tilde{T}_{2}\, ,
\,\,\,\,
P_{4}=-{\textstyle\frac{\sin{\alpha}}{R}}T_{1}
+{\textstyle\frac{\cos{\alpha}}{R}}\tilde{T}_{3}\, .
\end{equation}

The subalgebra $\mathfrak{h}$ is generated by $M$ and the orthogonal
subspace $\mathfrak{k}$ by the $P_{a}$s. The non-vanishing commutators

\begin{equation}
  \begin{array}{c}
[M,P_{0}] = \cos{\alpha}P_{1}\, ,
\,\,\,\, 
[M,P_{1}] = \cos{\alpha}P_{0}\, ,
\,\,\,\, 
[M,P_{2}] = -\sin{\alpha}P_{3}\, ,
\,\,\,\, 
[M,P_{3}]= \sin{\alpha}P_{2}\, ,
\\
\\
\left[P_{4},P_{0}\right]= -\frac{\sin{\alpha}}{R}P_{1}\, ,
\,\,\,\,
\left[P_{4},P_{1}\right]= -\frac{\sin{\alpha}}{R}P_{0}\, ,
\,\,\,\,
\left[P_{4},P_{2}\right]= -\frac{\cos{\alpha}}{R}P_{3}\, ,
\,\,\,\,
\left[P_{4},P_{3}\right]= -\frac{\cos{\alpha}}{R}P_{2}\, ,
\\
\\
\left[ P_{0}, P_{1}\right]= 
\frac{\cos{\alpha}}{R^{2}} M -\frac{\sin{\alpha}}{R}P_{4}\, ,
\,\,\,\,
[P_{2}, P_{3}]= 
-\frac{\sin{\alpha}}{R^{2}} M -\frac{\cos{\alpha}}{R}P_{4}\, ,
\\
\end{array}
\end{equation}

\noindent
indicate that $[\mathfrak{k},\mathfrak{h}]\subset \mathfrak{k}$
(reductivity) but $[\mathfrak{k},\mathfrak{k}]\not\!\subset
\mathfrak{h}$, so we do not have a symmetric pair and we will not have
a symmetric space.

The Killing metric of the product group manifold $SO(2,1)\times SO(3)$
in the new basis $\{P_{a},M\}$ $a=0,\cdots,4$ is

\begin{equation}
(K_{IJ})= 
\frac{2}{R^{2}}
\left(
  \begin{array}{cccccc}
-1 &    &    &    &    &    \\
   & +1 &    &    &    &    \\
   &    & -1 &    &    &    \\
   &    &    & -1 &    &    \\
   &    &    &    & 
\sin^{2}{\alpha}-\cos^{2}{\alpha} & -2R\sin{\alpha}\cos{\alpha}\\
   &    &    &    &    &    \\
   &    &    &    &    
-2R\sin{\alpha}\cos{\alpha}\,\,\,\,  & R^{2}(\cos^{2}{\alpha}-\sin^{2}{\alpha})
                            \\
  \end{array}
\right)\, ,
\end{equation}

\noindent
but we are not going to use it to construct the left-invariant metric.
Instead, we will use the 5-dimensional Minkowski metric $\eta_{ab}$,
which gives a metric invariant under the left action of $G$ since

\begin{equation}
f_{M(a}{}^{c}\eta_{b)c}=0\, .
\end{equation}

The coset representative is chosen to be

\begin{equation}
u(x)=e^{x^{0}P_{0}}\cdots e^{x^{4}P_{4}}\, ,  
\end{equation}

\noindent
and the left-invariant Maurer-Cartan 1-form $V=-u^{-1}du$
is

\begin{equation}
  \begin{array}{rcl}
-V & = & 
T_{I}\Gamma_{\rm Adj}(e^{-x^{4}P_{4}})^{I}{}_{J}
\Gamma_{\rm Adj}(e^{-x^{1}P_{1}})^{I}{}_{P_{0}}dx^{0}
+T_{I}\Gamma_{\rm Adj}(e^{-x^{4}P_{4}})^{I}{}_{P_{1}}dx^{1}\\
& & \\
& &  
+T_{I}\Gamma_{\rm Adj}(e^{-x^{4}P_{4}})^{I}{}_{J}
\Gamma_{\rm Adj}(e^{-x^{3}P_{3}})^{I}{}_{P_{2}}dx^{2}
+T_{I}\Gamma_{\rm Adj}(e^{-x^{4}P_{4}})^{I}{}_{P_{3}}dx^{3}
+P_{4}dx^{4}\, ,\\
\end{array}
\end{equation}

\noindent
and, with the definitions

\begin{equation}
V=e^{a}P_{a}+\vartheta M\, ,  
\end{equation}

\noindent
leads to the F\"unfbeins $e^{a}$ and to the $H$-connection $\vartheta$

\begin{equation}
  \begin{array}{rcl}
-e^{0} & = & 
{\rm cosh}\, (\frac{x^{1}}{R})\, 
{\rm cosh}\, (\frac{\sin{\alpha}}{R}x^{4})\, dx^{0}
+{\rm sinh}\, (\frac{\sin{\alpha}}{R}x^{4})\, dx^{1}\, ,\\
& & \\
-e^{1} & = & 
{\rm cosh}\, (\frac{x^{1}}{R})\, 
{\rm sinh}\, (\frac{\sin{\alpha}}{R}x^{4})\, dx^{0}
+{\rm cosh}\, (\frac{\sin{\alpha}}{R}x^{4})\, dx^{1}\, ,\\
& & \\
-e^{2} & = & 
\cos{\frac{x^{3}}{R}}\cos{(\frac{\cos{\alpha}}{R}x^{4})}\, dx^{2}
-\sin{(\frac{\cos{\alpha}}{R}x^{4})}\, dx^{3}\, ,\\
& & \\
-e^{3} & = & 
\cos{\frac{x^{3}}{R}}\sin{(\frac{\cos{\alpha}}{R}x^{4})}\, dx^{2}
+\cos{(\frac{\cos{\alpha}}{R}x^{4})}\, dx^{3}\, ,\\
& & \\
-e^{4} & = & 
-\sin{\alpha}\, {\rm sinh}(\frac{x^{1}}{R})\, dx^{0}
-\cos{\alpha}\sin{\frac{x^{3}}{R}}dx^{2}+dx^{4}\, ,\\
& & \\
-\vartheta & = & 
\frac{\cos{\alpha}}{R}\, {\rm sinh}(\frac{x^{1}}{R})\, dx^{0}
-\sin{\alpha}\sin{\frac{x^{3}}{R}}dx^{2}\, .\\
  \end{array}
\end{equation}

Redefining the coordinates

\begin{equation}
x^{0}/R = \phi\, ,
\,\,\,\,
x^{1}/R = \chi\, ,
\,\,\,\,
x^{2}/R = \varphi\, ,
\,\,\,\,
x^{3}/R = \theta +\pi/2\, ,
\,\,\,\,
x^{4}/R=\psi\, ,
\end{equation}

\noindent
it is easy to see that the metric

\begin{equation}
ds^{2}= \eta_{ab}e^{a}\otimes e^{b}\, ,  
\end{equation}

\noindent
is precisely that of Eq.~(\ref{eq:solution}).

According to the general results on homogeneous spaces the Killing
vectors $k_{(I)}$ associated to the left isometry group
$G=SO(2,1)\times SO(3)$ are given by

\begin{equation}
k_{(I)}=\Gamma_{\rm Adj}(u^{-1})^{a}{}_{I} e_{a}\, .
\end{equation}

Their explicit expressions are

\begin{equation}
  \begin{array}{rcl}
k_{(P_{0})} & = & 
-\partial_{x^{0}}\, ,\\
& & \\
k_{(P_{1})} & = & 
{\rm tgh}(x^{1}/R) \sin{(x^{0}/R)}\partial_{x^{0}}
-\cos{(x^{0}/R)}\partial_{x^{1}} 
-\sin{\alpha}{\displaystyle\frac{\sin{(x^{0}/R)}}{{\rm cosh}(x^{1}/R)}}
\partial_{x^{4}}\, ,\\
& & \\
k_{(P_{2})} & = & -\partial_{x^{2}}\, ,\\
& & \\
k_{(P_{3})} & = & 
-\tan(x^{3}/R) \sin{(x^{2}/R)}\partial_{x^{2}}
-\cos{(x^{2}/R)}\partial_{x^{3}} 
-\cos{\alpha}{\displaystyle\frac{\sin{(x^{2}/R)}}{\cos{(x^{3}/R)}}}
\partial_{x^{4}}\, ,\\
\end{array}
\end{equation}

\begin{equation}
  \begin{array}{rcl}
k_{(P_{4})} & = & 
\sin{\alpha}\left[{\rm tgh}(x^{1}/R)\cos{(x^{0}/R)}\partial_{x^{0}}
+\sin{(x^{0}/R)}\partial_{x^{1}}\right]\\
& & \\
& & 
-\cos{\alpha} \left[\tan{(x^{3}/R)}\cos{(x^{2}/R)}\partial_{x^{2}}
-\sin{(x^{2}/R)}\partial_{x^{3}} \right]\\
& & \\
& & 
-\left\{
{\displaystyle\frac{\cos{(x^{0}/R)}}{{\rm cosh}(x^{1}/R)}}
-\cos^{2}{\alpha}
\left[
{\displaystyle\frac{\cos{(x^{2}/R)}}{\cos{(x^{3}/R)}}}
-{\displaystyle\frac{\cos{(x^{0}/R)}}{{\rm cosh}{(x^{1}/R)}}}
\right] 
\right\}\partial_{x^{4}}\, ,\\
& & \\
k_{(M)} & = & 
-R\cos{\alpha}\left[{\rm tgh}(x^{1}/R)\cos{(x^{0}/R)}\partial_{x^{0}}
+\sin{(x^{0}/R)}\partial_{x^{1}}\right]\\
& & \\
& & 
-R\sin{\alpha} \left[\tan{(x^{3}/R)}\cos{(x^{2}/R)}\partial_{x^{2}}
-\sin{(x^{2}/R)}\partial_{x^{3}} \right]\\
& & \\
& & 
-R\sin{\alpha}\cos{\alpha}
\left[
{\displaystyle\frac{\cos{(x^{2}/R)}}{\cos{(x^{3}/R)}}}
-{\displaystyle\frac{\cos{(x^{0}/R)}}{{\rm cosh}{(x^{1}/R)}}}
\right] \partial_{x^{4}}\, .\\
\end{array}
\end{equation}

The right isometry group is given by the vectors dual to the
Maurer-Cartan 1-forms $e_{a}$ associated to the generators of
$N(H)/H$, and commute with the left Killing vectors. In this case, the
generator of $N(H)/H$ is $P_{4}$ and the associated Killing vector
denoted $k_{(N)}$ turns out to be

\begin{equation}
k_{(N)} = e_{4}= -\partial_{x^{4}}\, . 
\end{equation}


\section*{Construction of the  Killing  Spinors and the Superalgebra}

The Killing spinor equation of $N=2,d=5$ Supergravity is (choosing
$s(\alpha)=+1$) \cite{Cremmer:1980gs,Lozano-Tellechea:2002pn}

\begin{equation}
\left\{ \nabla_{a}
-{\textstyle\frac{1}{8\sqrt{3}}}
(\gamma^{bc}\gamma_{a}+2\gamma^{b}g^{c}{}_{a})
\mathcal{F}_{bc}\right\}\kappa=0\, .
\end{equation}

\noindent
$\kappa$ is an unconstrained Dirac spinor (one component of a pair of
symplectic-Majorana spinors).  We contract this equation with the
Maurer-Cartan 1-forms $e^{a}$ to write is in the form:

\begin{equation}
\left\{d -{\textstyle\frac{1}{4}}\omega^{a}{}_{b}\gamma_{a}{}^{b}  
-{\textstyle\frac{1}{8\sqrt{3}}}
(\gamma^{bc}\mathcal{F}_{bc}
\gamma_{a}+2\gamma^{b}\mathcal{F}_{ba})e^{a}
\right\}\kappa=0\, .
\end{equation}

In homogeneous spaces, the spin connection is given by

\begin{equation}
\omega^{a}{}_{b} = \vartheta^{i}f_{ib}{}^{a} 
+{\textstyle\frac{1}{2}} e^{c}f_{cb}{}^{a}\, , 
\end{equation}

\noindent
and we obtain a spinorial representation of the vertical generators $M_{i}$

\begin{equation}
\Gamma_{s}(M_{i})= {\textstyle\frac{1}{4}}f_{ib}{}^{a} 
\gamma_{a}{}^{b}\, .   
\end{equation}

In symmetric spaces the structure constants $f_{cb}{}^{a}=0$ and the
contribution of the spin connection to the Killing spinor equation is just
$-\vartheta^{i}\Gamma_{s}(M_{i})$ \cite{Alonso-Alberca:2002gh}, but in this
case we have extra terms

\begin{equation}
-{\textstyle\frac{1}{4}}\omega^{a}{}_{b} \gamma_{a}{}^{b}= 
-\vartheta\Gamma_{s}(M)
-{\textstyle\frac{1}{8}} e^{c}f_{cb}{}^{a} \gamma_{a}{}^{b}\, ,
\hspace{1cm}
\Gamma_{s}(M) \equiv {\textstyle\frac{1}{2}}
(\cos{\alpha} \gamma^{01} -\sin{\alpha}\gamma^{23})\, .
\end{equation}

\noindent
The extra terms do not give by itself $-e^{a}\Gamma_{s}(P_{a})$, but
it can be checked that, combined with the terms that depend on the
vector field strength, they do, and the Killing spinor equation take
the form

\begin{equation}
\left\{d -\vartheta\Gamma_{s}(M)  -e^{a}\Gamma_{s}(P_{a})\right\}\kappa=0\, ,
\hspace{1cm}
\Gamma_{s}(P_{a}) = -{\textstyle\frac{1}{2R}}
(\cos{\alpha} \gamma^{01} -\sin{\alpha}\gamma^{23}) \gamma_{a}\, ,
\end{equation}

\noindent
which leads to

\begin{equation}
\kappa = \Gamma_{s}(u^{-1})\kappa_{0}\, ,
\end{equation}

\noindent
where $\kappa_{0}$ is a constant spinor. The matrix
$\Gamma_{s}(u^{-1})^{\beta}{}_{\alpha}$ can be used as a basis of
Killing spinors $\kappa_{(\alpha)}{}^{\beta}$ to which we associate
supercharges $Q_{(\alpha)}$.

The commutators of the bosonic generators $P_{a},M,N$ of the
superalgebra (associated to the Killing vectors) with the supercharges
is given immediately by the spinorial Lie-Lorentz derivative of the
Killing spinor with respect to the associated Killing spinors
\cite{Figueroa-O'Farrill:1999va,Ortin:2002qb}. For the generators
associated to the left isometry group $\{T_{I}\}=\{P_{a},M\}$ we can
use Eq.~(2.23) of Ref.~\cite{Alonso-Alberca:2002gh}

\begin{equation}
\mathbb{L}_{k_{(I)}} \Gamma_{s}(u^{-1}) =
-\Gamma_{s}(u^{-1}) 
[\mathbb{L}_{k_{(I)}} \Gamma_{s}(u)] \Gamma_{s}(u^{-1}) = 
-\Gamma_{s}(u^{-1}) \Gamma_{s}(T_{I})\, ,
\end{equation}

\noindent 
which implies the commutators

\begin{equation}
\label{eq:QTcommutators}
[Q_{(\alpha)},T_{I}] = -Q_{(\beta)}  \Gamma_{s}(T_{I})^{\beta}{}_{\alpha}\, .
\end{equation}

The other commutators with $N$ are trivial.

Finally, let us consider the anticommutators of two supercharges.
These are associated to the decomposition in Killing vectors the
bilinears $-i\bar{\kappa}_{\alpha}\gamma^{a}\kappa_{(\beta)}e_{a}$.
To find this decomposition is crucial to relate the contravariant
gamma matrices $\gamma^{a}$ with the bosonic generators in the
spinorial representation $\Gamma_{s}(P_{a})$. In this case, it is
convenient to proceed as follows. First, we find the relation

\begin{equation}
\gamma^{a}= \eta^{ab}\gamma_{b}=-2R\mathcal{S}\Gamma_{s}(P_{a})\, ,
\hspace{1cm}
\mathcal{S}= (\cos{\alpha} \gamma^{01} +\sin{\alpha}\gamma^{23})\, ,   
\end{equation}

\noindent 
and substitute into the bilinear 

\begin{equation}
-i\bar{\kappa}_{(\alpha)}\gamma^{a}\kappa_{(\beta)}e_{a} = 
-i\Gamma_{s}(u^{-1})^{\dagger}\mathcal{D}\mathcal{S}\Gamma_{s}(P_{b})
\Gamma_{s}(u^{-1})\eta^{ba}e_{a}\, ,  
\end{equation}

\noindent
where $\mathcal{D}=i\gamma^{0}$ is th Dirac conjugation matrix. It can
be checked that

\begin{equation}
\Gamma_{s}(u^{-1})^{\dagger}\mathcal{D}\mathcal{S} =
\mathcal{D}\mathcal{S}\Gamma_{s}(u)\, ,
\end{equation}

\noindent
and, recognizing the adjoint action of $u$ on the $\Gamma_{s}(P_{b})$
we have

\begin{equation}
-i\bar{\kappa}_{\alpha}\gamma^{a}\kappa_{(\beta)}e_{a}
=-i\mathcal{D}\mathcal{S}\Gamma_{s}(T_{I})\Gamma_{\rm Adj}(u)^{I}{}_{b}
\eta^{ba}e_{a}\, .   
\end{equation}

\noindent
Now we use the following general property: for any $g\in G$, (if the
Killing metric is nonsingular, as here)

\begin{equation}
\Gamma_{\rm Adj}(g)^{I}{}_{J} = 
K_{JK}\Gamma_{\rm Adj}(g^{-1})^{K}{}_{L} K^{LI}\, ,  
\end{equation}

\noindent
and the definition of the dual generators $T^{I}= K^{IJ}T_{J}$

\begin{equation}
-i\bar{\kappa}_{\alpha}\gamma^{a}\kappa_{(\beta)}e_{a}
=-i\mathcal{D}\mathcal{S}\Gamma_{s}(T^{I})
\Gamma_{\rm Adj}(u^{-1})^{J}{}_{I}K_{Jb}\eta^{ba}e_{a}\, .   
\end{equation}

Since the Killing metric and the Minkowski metric are different, the
r.h.s.~of this expression does not give the Killing vectors of the
left isometry group. We have to use a non-trivial property of
$\Gamma_{s}(u^{-1})$. Let us define the matrix $\eta^{IJ}$

\begin{equation}
(\eta^{IJ})
= 
\left( 
\begin {array}{cccccc} 
1& & & & & \\
 &-1& & & & \\
 & &-1& & & \\
 & & &-1& & \\
 & & & &-1& \\
 & & & & &-{R}^{-2}
\end {array} \right)\, ,
 \end{equation}

\noindent
and, with it and the Killing metric, the matrix 

\begin{equation}
R^{I}{}_{J}=\eta^{IK}K_{KJ}\, .  
\end{equation}
 
\noindent 
It can be checked that 

\begin{equation}
R^{I}{}_{J}\Gamma_{\rm Adj}(u^{-1})^{J}{}_{K}= 
\Gamma_{\rm Adj}(u^{-1})^{I}{}_{L} R^{L}{}_{K}\, ,
\Rightarrow
\Gamma_{\rm Adj}(u^{-1})^{J}{}_{I}K_{Jb}\eta^{ba}= 
\Gamma_{\rm Adj}(u^{-1})^{a}{}_{L}R^{L}{}_{I}\, ,
\end{equation}

\noindent 
and

\begin{equation}
-i\bar{\kappa}_{\alpha}\gamma^{a}\kappa_{(\beta)}e_{a}
=-i\left[\mathcal{D}\mathcal{S}\Gamma_{s}(T^{I})\right]_{\alpha\beta} 
R^{L}{}_{I} k_{(L)}\, ,
\end{equation}

\noindent
that gives the anticommutators

\begin{equation}
\{Q_{(\alpha)},Q_{(\beta)}\}=
-i\left[\mathcal{D}\mathcal{S}\Gamma_{s}(T^{I})\right]_{\alpha\beta} 
(R^{a}{}_{I} P_{a} + R^{M}{}_{I}M)\, .
\end{equation}


\section*{Acknowledgments}

T.O.~would like to thank P.~Meessen for most useful conversations and
M.M.~Fern\'andez for her continuous support.  This work has been
partially supported by the Spanish grant FPA2000-1584.

\appendix





\end{document}